\documentclass[letterpaper]{article}
\usepackage{authblk}
\usepackage{url}
\usepackage[margin=1in,footskip=0.25in]{geometry}
\usepackage[labelfont=bf]{caption}
\usepackage{graphicx}
\usepackage{amsmath,amssymb,amsfonts,amsmath}
\usepackage{graphicx}
\usepackage{subfigure}
\usepackage{caption}
\usepackage{tikz}
\usepackage{float}
\usepackage[dvipdf]{epsfig}
\usepackage{float}
\usepackage{lipsum}
\usepackage{wrapfig}
\usepackage{cite}
\usepackage{color}
\usepackage{easylist}
\usepackage{lineno}
\usepackage{booktabs}
\usepackage{hyperref}
\usepackage{cleveref}
\usepackage{multirow}
\usepackage{hhline}
\usepackage{soul}
\usepackage{blindtext}
\usepackage{threeparttable}
\usepackage[bbgreekl]{mathbbol}

\usepackage{graphicx}
\graphicspath{{./Images/}}
\usepackage{dcolumn}
\usepackage{bm}

\usepackage[utf8]{inputenc}
\usepackage[T1]{fontenc}
\usepackage{mathptmx}
\usepackage{etoolbox}
\usepackage{subfigure}
\usepackage{hyperref}
\usepackage{color}
\usepackage{amsmath,amssymb,amsfonts,amsmath}
\usepackage{textcase}

\newcommand{\hbarr}{\mathchar'26\mkern-9mu h}

\newcommand{\eqn}[1]{
\begin{eqnarray}
#1
\end{eqnarray}
}

\newcommand{\tb}[1]{
\textbf{#1}
}

\newcommand{\ip}[1]{
\left\langle #1 \right\rangle
}

\newcommand{\pr}{
\partial_{r}
}

\newcommand{\px}{
\partial_{x}
}

\newcommand{\pz}{
\partial_{z}
}

\makeatletter
\def\@email#1#2{%
\endgroup
\patchcmd{\titleblock@produce}
{\frontmatter@RRAPformat}
{\frontmatter@RRAPformat{\produce@RRAP{*#1\href{mailto:#2}{#2}}}\frontmatter@RRAPformat}
{}{}
}%
\makeatother

\title{Vibrational Levels of a Generalized Morse Potential}

\author[1]{Saad Qadeer} 
\author[2]{Garrett D. Santis} 
\author[3]{Panos Stinis} 
\author[4]{Sotiris S. Xantheas\thanks{{\tt sotiris.xantheas@pnnl.gov} and \tt{xantheas@uw.edu}}} 
\affil[1]{Advanced Computing, Mathematics and Data Division, Pacific Northwest National Laboratory, 902 Battelle Boulevard, P.O. Box 999, MS K1-83, Richland, WA 99352, USA}
\affil[2]{Department of Chemistry, University of Washington, Seattle, WA 98195}
\affil[3]{Advanced Computing, Mathematics and Data Division, Pacific Northwest National Laboratory, 902 Battelle Boulevard, P.O. Box 999, MS K1-83, Richland, WA 99352, USA}
\affil[ ]{Department of Applied Mathematics, University of Washington, Seattle, WA 98195}
\affil[4]{Advanced Computing, Mathematics and Data Division, Pacific Northwest National Laboratory, 902 Battelle Boulevard, P.O. Box 999, MS K1-83, Richland, WA 99352, USA}
\affil[ ]{Department of Chemistry, University of Washington, Seattle, WA 98195}

\begin{document}
	
\maketitle

\begin{abstract}
A Generalized Morse Potential (GMP) is an extension of the Morse Potential (MP) with an additional exponential term and an additional parameter that compensate for MP's erroneous behavior in the long range part of the interaction potential. Because of the additional term and parameter, the vibrational levels of the GMP cannot be solved analytically, unlike the case for the MP. We present several numerical approaches for solving the vibrational problem of the GMP based on Galerkin methods, namely the Laguerre Polynomial Method (LPM), the Symmetrized Laguerre Polynomial Method (SLPM) and the Polynomial Expansion method (PEM) and apply them to the vibrational levels of the homonuclear diatomic molecules B$_2$, O$_2$ and F$_2$, for which high level theoretical Full CI potential energy surfaces and experimentally measured vibrational levels have been reported. Overall the LPM produces vibrational states for the GMP that are converged to within spectroscopic accuracy of 0.01 cm$^{-1}$ in between 1 and 2 orders of magnitude faster and with much fewer basis functions / grid points than the Colbert-Miller Discrete Variable Representation (CM-DVR) method for the three homonuclear diatomic molecules examined in this study. A python library that fits and solves the GMP and similar potentials can be downloaded from \url{https://gitlab.com/gds001uw/generalized-morse-solver}.
\end{abstract}

\section{Introduction}\label{SecIntro}

Since its introduction in 1929, the Morse Potential (MP)\cite{morse1929diatomic}
\eqn{
V_{\text{M}}(r)=D_{e}\left[1+e^{-2\alpha(r-r_{m})}-2e^{-\alpha(r-r_{m})}\right],
\label{eqMP}
}
has become the textbook example for describing the potential energy surface (PES) of interatomic interactions because of its simple form and the ability to produce analytical solutions for the vibrational energy levels, 
\eqn{E_{n}=h\nu _{0}(n+1/2)-{\frac {\left[h\nu _{0}(n+1/2)\right]^{2}}{4D_{e}}}.}
In the above $D_{e}$ is the dissociation energy, $r_{m}$ the distance at the minimum, $\alpha={\sqrt {k_{e}/2D_{e}}}$ with $k_{e}$ the force constant, $\nu_{0}={\frac {\alpha}{2\pi }}{\sqrt {2D_{e}/m}}$ with $m$ the mass of the particle and $n$ an integer corresponding to the vibrational quantum number. Note that the spacing of the vibrational levels $E_{n+1}-E_{n}=h\nu _{0}-(n+1)(h\nu _{0})^{2}/2D_{e}$ becomes zero for $n^{*}={(\frac {2D_{e}}{h\nu _{0}}-1)}$
and negative above it, indicating the inability of the MP to describe vibrational levels above $n^{*}$. For the systems that we will investigate in this work (\textit{vide infra}) $n^*$ is the last vibrational level below dissociation.

The simple form and limited flexibility of MP's potential function \ref{eqMP} also result in deviations from the true PES at distances larger than $r_{m}$. To alleviate this problem, more complicated extensions of the MP, such as the Morse/long-range (MLR) potential, have been previously introduced.\cite{LeRoy1, LeRoy2, LeRoy3, LeRoy4, LeRoy5} However, these typically include many more parameters and the vibrational levels are obtained using numerical methods. We have previously introduced a Generalized Morse Potential (GMP)\cite{xantheas2014universal}
\eqn{V_{\text{gM}}^{(\gamma)}(r)=D_{e}\left[1+\frac {\gamma}{\alpha}\left(e^{-2\alpha(r-r_{m})}-e^{-\alpha(r-r_{m})}\right)-e^{-\gamma(r-r_{m})}\right],
\label{eqGMP}
}
which is the simplest expansion of the original MP as it includes just one additional exponential term and one additional parameter $\gamma$, and reverts back to the original MP for $\gamma=\alpha$ and for $\gamma=2\alpha$. In this respect, the MP can be considered as a special case of the GMP with $\gamma=\alpha$ or $\gamma=2\alpha$. The GMP has been shown to produce superior fits to the intermolecular PESs of alkali metal - water cationic and halide - water anionic clusters in both the repulsive short range and attractive long range parts of the PES. \cite{werhahn2014universal} The GMP in terms of the reduced coordinates $r^*=r/r_m$, $\alpha^*=\alpha r_m$, $\gamma^*=\gamma r_m$ and $\epsilon^*=(V_{\text{gM}}^{(\gamma)}/D_e)-1$ can be cast as as\cite{xantheas2014universal}
\eqn{{\epsilon^*}=\frac{\gamma^*}{\alpha^*}\left(e^{2\alpha^*(1-r^*)}-e^{\alpha^*(1-r^*)}\right)-e^{\gamma^*(1-r^*)}.
\label{eqGMPred}
}
In this study we examine the solution of the vibrational levels of the GMP based on Galerkin methods. We compare the time--to-solution of three different approaches using Laguerre and Legendre polynomials compared to the standard Colbert-Miller Discrete Value Representation (DVR) methods\cite{CMDVR} when applying it to the PESs of the diatomic molecules $\text{B}_{2}$, $\text{O}_{2}$ and $\text{F}_{2}$, for which Full Configuration Interaction (FCI) PESs and experimentally measured vibrational levels up to the dissociation limit have been reported.

\section{The Laguerre Polynomial Method (LPM)}\label{SecLagExp}

The Schr\"{o}dinger equation for the GMP is
\eqn{
\left(-\frac{\hbarr^2}{2m}\pr^2 + V_{\text{gM}}^{(\gamma)}(r)\right)\psi(r) = E\psi(r), \label{Schrod}
}
where $V_{\text{gM}}^{(\gamma)}$ is the potential function for the GMP (\ref{eqGMP}). Defining $x = \alpha r$, $x_m = \alpha r_m$, and $\eta = \gamma/\alpha$, and setting $\hat V^{(\eta)}_{\text{gM}}(x) = \frac{2m}{\hbarr^2 \alpha^2}\left(V^{(\eta\alpha)}_{\text{gM}}(x/\alpha) - D_e\right)$ allows us to rewrite (\ref{Schrod}) as 
\eqn{
\left(-\px^2 + \hat V^{(\eta)}_{\text{gM}}(x)\right)\hat{\psi}(x) = \epsilon \hat{\psi}(x). \label{ScaSchrod}
}

Here, $\hat{\psi}(x) = \psi(x/\alpha)$ and $\epsilon = \frac{2m}{\hbarr^2 \alpha^2}(E-D_e)$. In addition, setting $\lambda^2 = \frac{2m}{\hbarr^2 \alpha^2}D_e$ gives
\eqn{
\hat V^{(\eta)}_{\text{gM}}(x) = \lambda^2\left(\eta\left(e^{-2(x-x_m)} - e^{-(x-x_m)}\right) - e^{-\eta(x-x_m)}\right). \label{ScaGM}
}

Finally, defining $z = 2\lambda e^{-(x-x_m)}$ and setting $\tilde {\psi}(z) = \hat{\psi}(x)$ yields
\eqn{
\left(z^2 \pz^2 + z\pz + q_{\eta}(z)\right)\tilde{\psi}(z) = -\epsilon \tilde{\psi}(z), \label{ScaGM2}
}
where 
\eqn{
q_{\eta}(z) = \frac{\lambda \eta}{2}z - \frac{\eta}{4}z^2 + \frac{\lambda^{2-\eta}}{2^{\eta}}z^{\eta}. \label{Qeta}
}

Observe that we recover the original MP for $\eta = 1$ and 2 since $q_1(z) = q_2(z) = \lambda z - \frac{z^2}{4}$. Following the approach by Morse \cite{morse1929diatomic}, the problem can be solved by decomposing $\tilde{\psi}(z) = e^{-z/2}z^{\beta/2}\phi(z)$ to obtain
\eqn{
z\phi''(z) + (\beta+1-z)\phi'(z) + \left(\lambda - \frac{(\beta+1)}{2}\right)\phi(z) + 
\left(\epsilon + \frac{\beta^2}{4}\right)\frac{\phi(z)}{z} = 0. \label{LagForm} 
}

Choosing $\epsilon = -\beta^2/4$ and setting $\beta = 2\lambda - 2n - 1 $ so that
\eqn{
\lambda - \frac{(\beta+1)}{2} = n, \nonumber
}
for any nonnegative integer $n$, leads to the generalized Laguerre differential equation
\eqn{
z\phi''(z) + (2\lambda-2n-z)\phi'(z) + n\phi(z) = 0. \label{LagEqn}
}

Bounded solutions of this equation are known as the generalized Laguerre polynomials $L^{(2\lambda-2n-1)}_n(z)$. This shows that the eigenvalues of (\ref{ScaGM2}) are $\epsilon_n = -(\lambda-n-1/2)^2$ with eigenfunctions 
\eqn{
\tilde{\psi}_n(z) = e^{-z/2}z^{\lambda-n-1/2}L_n^{(2\lambda-2n-1)}(z) \nonumber
} 
for $0 \leq n \leq \lfloor \lambda-1/2 \rfloor$.

For values of $\eta$ other than 1 or 2, it may not be possible to obtain analytical solutions of (\ref{ScaGM2}) so we resort to numerical methods for the general case. Broadly speaking, our techniques rely on expanding the unknown eigenfunctions in terms of suitably chosen bases and applying Galerkin conditions to reduce (\ref{ScaGM2}) to matrix eigenvalue problems. The three approaches we consider tackle the problem in subtly different ways. However, we find (\textit{vide infra}) that the resulting solutions agree with each other to a very high degree of precision, thus validating each other. We will only present the first method in the main part of the paper, with the other 2 methods outlined in Appendices \ref{SecSymm} and \ref{SecFinite}.

Define the inner products
\eqn{
\ip{f_1,f_2}_{L^2} = \int_{0}^{\infty} \overline{f_1(z)}f_2(z) \ dz, \label{InnProd}
}
and, for $\beta \geq 0$, 
\eqn{
\ip{g_1,g_2}_{\beta} = \int_{0}^{\infty} \overline{g_1(z)}g_2(z) e^{-z}z^{\beta} \ dz. \label{InnProd2}
}

Note that
\eqn{
\ip{h_1e^{-z/2}z^{\beta/2},h_2e^{-z/2}z^{\beta/2}}_{L^2} = \ip{h_1,h_2}_{\beta}. \label{IPconv}
}

The decomposition $\tilde{\psi}(z) = e^{-z/2}z^{\beta/2}\phi(z)$ used for the original MP suggests the use of basis functions of the form
\eqn{
u_j^{(\beta)}(z) = e^{-z/2}z^{\beta/2}p_j(z), \quad j \geq 0, \label{BasisFuncs}
} 
where $\{p_j\}_{j \geq 0}$ is a suitable family of polynomials. Generally, the assumption underlying Galerkin methods is that the solution to a problem can be well-approximated by a finite collection of functions. In our case, the basis functions (\ref{BasisFuncs}) are of the same form as the eigenfunctions for the original MP so these might be expected to serve as the building blocks for the generalized problem. The Galerkin conditions then require that the residual of this finite-dimensional approximation be orthogonal to the basis functions with respect to a suitable inner product. We impose the constraints  
\eqn{
\ip{u_l^{\beta},\left(z^2 \pz^2 + z\pz + q_{\eta}\right)\tilde{\psi}}_{L^2} = -\epsilon \ip{u_l^{(\beta)},\tilde{\psi}}_{L^2}, \label{GalCond}
}
for $l \geq 0$. We have 
\eqn{
\left(z^2 \pz^2 + z\pz + q_{\eta}(z)\right)u_j^{(\beta)}(z) 
&=&  e^{-z/2}z^{\beta/2}\left(z^2 p_j''(z) + z(\beta+1-z)p_j'(z) + \right. \nonumber\\
&& \hspace{0.2cm}\left.\left(\frac{z^2}{4} - \frac{(\beta+1)}{2}z + \frac{\beta^2}{4} + q_{\eta}(z)\right)p_j(z)\right). \label{BFPlug}
}

Requiring $\{p_j\}$ to be the orthonormalized generalized Laguerre polynomials that solve
\eqn{
z^2 p_j''(z) + z(\beta+1-z)p_j'(z) = -jzp_j(z) \label{LagEqn2}
}
for $j \geq 0$ then yields 
\eqn{
 \left(z^2 \pz^2 + z\pz + q_{\eta}(z)\right)u_j^{(\beta)}(z) &=&  e^{-z/2}z^{\beta/2}\left(\frac{z^2}{4} - \frac{(\beta+1 + 2j)}{2}z + \frac{\beta^2}{4} + q_{\eta}(z)\right)p_j(z). \nonumber\\ \label{BFPlug2}
}

Thus, substituting the expansion $\tilde{\psi}(z) = \sum_{j = 0}^M c_j u_j^{(\beta)}(z)=e^{-z/2}z^{\beta/2}\sum_{j = 0}^M c_j p_j(z)$ into (\ref{GalCond}) and using only $0 \leq l \leq M$ results in
\eqn{
\sum_{j = 0}^M \ip{p_l,\left(\frac{z^2}{4} - \frac{(\beta+1 + 2j)}{2}z + \frac{\beta^2}{4} + q_{\eta}\right)p_j}_{\beta} c_j = -\epsilon c_l, \nonumber\\ \label{EigProb}
}
where we have used (\ref{IPconv}) and the fact that the generalized Laguerre polynomials $\{p_j\}$ are orthonormal with respect to (\ref{InnProd2}). Observe that with this choice for the $\{p_j\}$, we do not need to compute the derivatives of the polynomials as they show up in (\ref{BFPlug}). Equation (\ref{EigProb}) can be recognized as an eigenvalue problem for the $(M+1) \times (M+1)$ matrix $S^{(\eta)}$ defined by
\eqn{
S^{(\eta)}_{lj} = \ip{p_l,\left(\frac{z^2}{4} - \frac{(\beta+1 + 2j)}{2}z + \frac{\beta^2}{4} + q_{\eta}\right)p_j}_{\beta}, \label{SMat} 
}
for $0\leq l,j \leq M$. We note that for integer values of $\eta \geq 1$, the orthogonality of the generalized Laguerre polynomials and the fact that 
\eqn{
\text{deg}\left(\frac{z^2}{4} - \frac{(\beta+1 + 2j)}{2}z + q_{\eta}(z)\right) \leq \eta \nonumber
}
lead to $S_{lj}^{(\eta)} = 0$ for $|l-j| > \eta$. (For $\eta = 2$, it reduces to the $\eta = 1$ tri-diagonal matrix due to the cancellation of the $z^2/4$ term.) However, due to the $(jzp_j)$ term, it is guaranteed to be non-symmetric. The non-trivial entries of this matrix can be computed by employing Gauss--Laguerre quadrature. This needs to be of sufficiently high-resolution, particularly in the cases where $\eta$ is a non-integer and the integrands are not polynomials. 

\section{Comparison between the various Galerkin based Methods for obtaining the vibrational levels}

The previous section outlines the most straightforward method to solve for the vibrations levels of the GMP. However, we also present two alternative techniques, termed the Symmetrized Laguerre Polynomial Method (SLPM, see Appendix \ref{SecSymm}) and the Polynomial Expansion Method (PEM, see Appendix \ref{SecFinite}).  Besides these three basis set methods, approaches
based on Discrete Variable Representation (DVR) have been previously reported and can be applied to a wide range of systems including the GMP.\cite{CMDVR,PODVR}  One of the most popular methods in that category is the Colbert-Miller DVR (CM-DVR),\cite{CMDVR} which uses localized and uniformly distributed basis functions to create a simple potential energy matrix (the potential function on the diagonal), and a more complex but analytically evaluated kinetic energy matrix.  
A python library and scripts contained in \url{https://gitlab.com/gds001uw/generalized-morse-solver} fit the GMP and similar potentials and solve for their vibrational levels up to dissociation using the methods (LPM, SLPM, PEM and CM-DVR) discussed in this study. In this section we evaluate the time--to--solution for the various methods and show that the Galerkin based methods using Laguerre Polynomials presented herein are significantly faster than the CM-DVR methods while retaining accuracy up to dissociation.

The three Galerkin based methods (cf. Section \ref{SecLagExp} and Appendices \ref{SecSymm} and \ref{SecFinite})
were tested on an arbitrary system to demonstrate the transferability of these methods. For this purpose, a GMP with $r_m=\sqrt2  \text{\AA} $, $D_e=0.577 E_h$, $\mu=1.618$ amu, $\alpha=\pi$, $\eta = \frac{\gamma}{\alpha} = \ln{10}$, and $\lambda=46.200$ was used. Note that $\lambda$ is a good metric for the number of eigenstates that exist before dissociation. The reference eigenstates of this potential were obtained via CM-DVR in the range $r^* \in [0.4 ,25]$ using 10,000 grid points. This was found to be a large enough range of values for $r^*$ to converge to the final eigenstate and an adequate number of grid-points to converge the all energies. The performance of each method was assessed by increasing the number of basis functions till convergence or the onset of numerical instabilities. The value of $beta$ was zero for the LPM simulations and the one for the SLPM. CM-DVR calculations used to test convergences were performed in the range $r^* \in [0.5,5]$. Errors were computed by comparing the eigenvalues from each method against those from the reference CM-DVR calculation. Timings were collected as an average of numerous runs that included setting up the solution matrix and diagonalization, but did not include the step of initializing the potential.  Times are reported for an 8 core AMD Razen 7 1700X processor (2200-3400 MHz) with the \texttt{scipy.linalg} library for algebraic operations.\cite{2020SciPy-NMeth}

\begin{figure}[tbph]
\centering
\includegraphics[width=\textwidth]{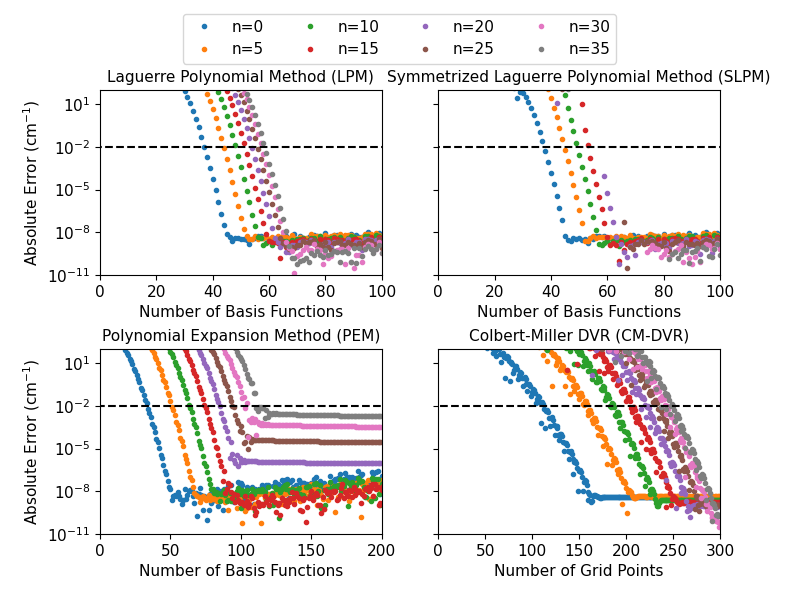}
\caption{The convergence plots of the lower eigenstates of the GMP are shown for the Laguerre Polynomial Method (LPM), Symmetrized Laguerre Polynomial Method (SLPM, Appendix \ref{SecSymm}), the Polynomial Expansion Method (PEM, Appendix \ref{SecFinite}), and for Colbert-Miller Discrete Variable Representation (CM-DVR \cite{CMDVR}).}
\label{fig:lowConverge}
\end{figure}

\begin{figure}[tbph]
\centering
\includegraphics[width=1.0\textwidth]{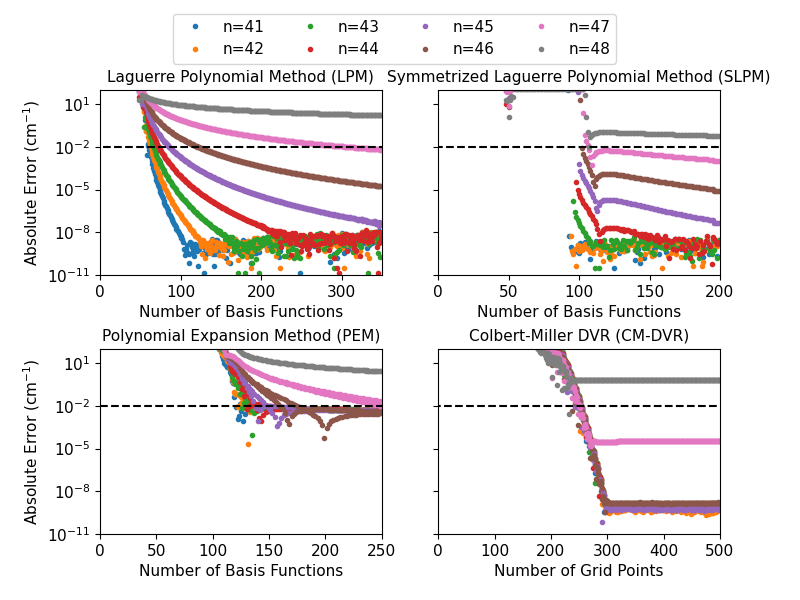}
\caption{The convergence plots of the last few (highest) eigenstates of the GMP are shown for the Laguerre Polynomial Method (LPM), Symmetrized Laguerre Polynomial Method (SLPM, Appendix \ref{SecSymm}), the Polynomial Expansion Method (PEM, Appendix \ref{SecFinite}), and for the Colbert-Miller Discrete Variable Representation (CM-DVR) \cite{CMDVR}. These states are harder to approximate through the use of basis functions based on solutions of the MP.}
\label{fig:highConverge}
\end{figure}

\begin{figure}[tbph]
\centering
\includegraphics[width=1.0\textwidth]{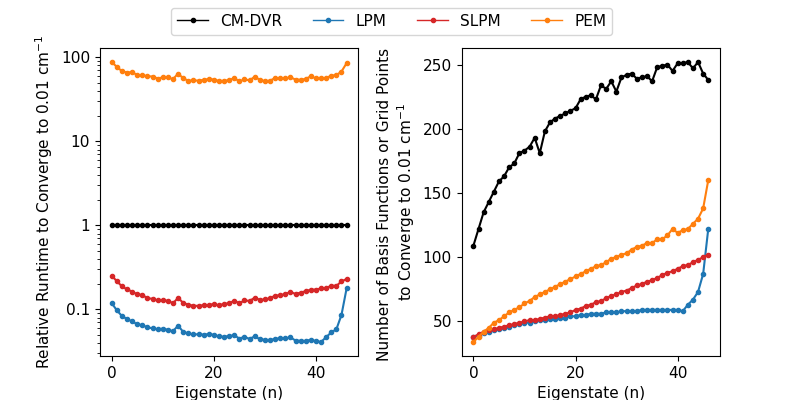}
\caption{The convergence of the GMP summarized for the Laguerre Polynomial Method (LPM), Symmetrized Laguerre Polynomial Method (SLPM, Appendix \ref{SecSymm}), the Polynomial Expansion Method (PEM, Appendix \ref{SecFinite}), and for the Colbert-Miller Discrete Variable Representation (CM-DVR) \cite{CMDVR}. (Left panel) time (relative to CM-DVR) required to run each converged simulation, which includes both assembling the matrix and the diagonalization. (Right panel): number of basis functions or grid points required to converge the eigenvalues to 0.01 cm$^{-1}$, which controls the size of the linear system solved for each method.}
\label{fig:timeConverge}
\end{figure}

The convergence of the lower eigenstates was studied first (Figure \ref{fig:lowConverge}).  Notably, all the eigenstates below $n=35$ converge relatively quickly within 0.01 cm$^{-1}$.  Vibrational states converge with a number of basis functions that lies between one and two multiples of $\lambda$ for the LPM and the SLPM.  For DVR, all eigenvalues converge with less than 300 grid points, which agrees with the method used by Bytautas \textit{et al.}, \cite{F2_III,B2,O2} who performed CM-DVR on chemical systems similar to that modeled by these parameters.  Eigenstates closer to dissociation behave differently as can be seen in Figure \ref{fig:highConverge}. Since these states are in regimes where the potential is less Morse-like, accurate representations of the eigenfunctions necessitates the use of more basis functions. However, all but the highest states converge within spectroscopic accuracy (0.01 cm$^{-1}$), further validating the LPM and SLPM methods for the vibrational states of importance. For each vibrational state there exists a minimum number of basis functions / grid points that are required to converge it to some threshold at some run time. Figure \ref{fig:timeConverge} indicates the number of basis functions or grid-points required to converge each vibrational state to 0.01 cm$^{-1}$ as well as the corresponding run-times.  The LPM and the SLPM are a full order of magnitude faster than the literature standard, CM-DVR.  In summary, the LPM is the fastest method for finding the eigenstates for the GMP.  The SLPM is comparable to the LPM, but suffers from numerical stability issues, and thus must be used with caution.  DVR can achieve the greatest accuracy because it is highly stable, but this comes at increased cost.  It is important, however, to recognize that CM-DVR is transferable to any potential surface but LPM and SLPM are specific to Morse-like potentials.

The convergence of the PEM is discussed separately due to its different behavior.  The method does converge for all eigenvalues, barring the last two, while employing 250 basis functions; other methods were also found to have problems in that regime.  However, the states do not converge to computer precision ($\sim 10^{-8}$), but only to spectroscopic accuracy ($\sim 10^{-2}$).  Furthermore, the time required to converge is high compared to other methods;  there is a 2 orders of magnitude slowdown for the PEM compared to CM-DVR and a 3 orders of magnitude slowdown compared to the LPM. The slowdown is a consequence of the polynomial basis being required to depict the rapid activity for low values of the rescaled independent variable $z$ as well as the fast decay for larger values. In contrast, the LPM and SLPM employ bases with built-in exponentially decaying factors and hence perform better with fewer basis functions.

\section{Application to the vibrational levels of the homonuclear diatomic molecules \texorpdfstring{B\textsubscript{2}, O\textsubscript{2} and F\textsubscript{2}}{ }}

\subsection{Fits of the PESs}
\begin{figure}[tbph]
\centering
\includegraphics[width=\textwidth]{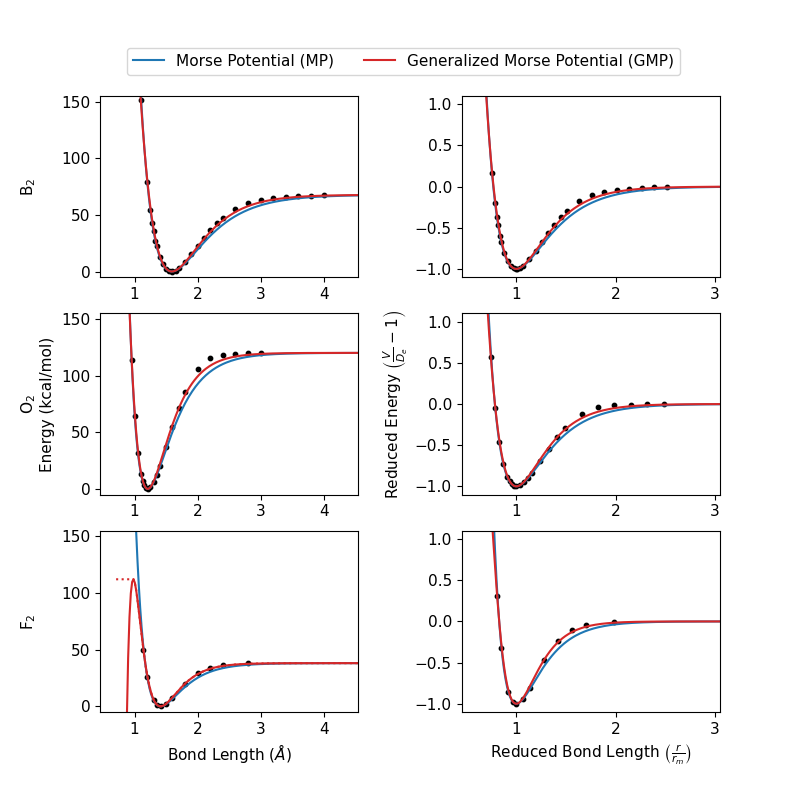}
\caption{Fits of the MP (red) and the GMP (blue) to the \textit{ab-initio} PESs of B$_2$, O$_2$, and F$_2$ reported by Bytautas \textit{et al.}\cite{B2,O2,F2_III} The dotted line for F$_2$ shows a necessary modification that can be applied to prevent the turnover for simulations or CM-DVR calculations.} 
\label{fig:Pots}
\end{figure}

The GMP has been previously used to fit the intermolecular PESs corresponding to charge--dipole electrostatic interactions in alkali metal and halide -- water clusters.\cite{werhahn2014universal} The extension to covalent bonds presented in this study is a natural choice since the MP is already a good and well used PES for covalent interactions.\cite{PeFuncForDiatomicMolecules,MpParamsForInteractions,ModelingOfDiatomicMolecule} However, it has been acknowledged that the MP does not fit covalent interactions well at larger interatomic distances (cf. Figure \ref{fig:Pots}), and therefore attempts have been made to improve such potentials.\cite{PeFuncForDiatomicMolecules,xantheas2014universal} The availability of highly accurate \textit{ab-initio} potential energy surfaces for O$_2$, F$_2$, and B$_2$ that include core and valence correlation and relativistic effects\cite{F2_II,O2,B2} presents a unique opportunity to quantitatively compare theoretically predicted molecular constants and properties such as the dissociation energy, the vibrational levels, or the vibrationally averaged equilibrium bond length to experimentally measured quantities. Bytautas \textit{et al.} have used their high level theoretical PESs for the above three diatomics in conjunction with a highly parameterized even-tempered Gaussian function (ETGP) to obtain vibrational levels in outstanding agreement with the measured ones up to dissociation.\cite{F2_III,O2,B2} In this work the \textit{ab-initio} PESs of O$_2$, F$_2$, and B$_2$ were analyzed based on the MP and GMP to gauge the performance of these much simpler functions with respect to the more complex and more accurate ETGP approach.

In this work, all fits and calculations were performed in reduced coordinates.\cite{xantheas2014universal} This produces comparable potentials and results that can be unscaled to get physical values with the appropriate units for each system.  Figure \ref{fig:Pots} shows the fits with the MP and GMP for the \textit{ab-initio} PESs of O$_2$, F$_2$, B$_2$ reported by Bytautas \textit{et al.}\cite{B2,O2,F2_III}. There is a clear improvement of the fit with the additional exponent in the GMP over the MP that is already even visually recognized.  The parameters of the fits are supplied in Tables \ref{tab:paramsUnits} and \ref{tab:paramsReduced}, in which the root mean square differences (RMSD) are evaluated between the analytic function(s) and the highly accurate \textit{ab-initio} points.  We find an approximate 50\% reduction in the error of the fit with the additional exponential (GMP over MP).  It is promising that minimal modifications to the MP can yield highly accurate PESs.  It is important to note that a large error still exists for O$_2$, the only diatomic in the set to have a bond order of two.  For that reason, the GMP may need additional enhancements to account for breaking a double bond.

\begin{table}[tbph]
\centering
\caption{Fitted parameters for the MP and GMP for B$_2$, O$_2$, and F$_2$.  The RMSD of the various potential fits and the \textit{ab-initio} PESs\cite{F2_II,O2,B2} are also indicated.\\}
\begin{tabular}{|c|c c|c| c c|c c c|}
\hline Molecule\  &\ $r_m$\ (\AA)\ &\ $D_e$\ (kcal/mol) &$\alpha$ (\AA$^{-1}$) & $\alpha$ (\AA$^{-1}$) &\ $\gamma$ (\AA$^{-1}$) \ &\multicolumn{3}{c|}{RMSD (kcal/mol)}\\\hline
&  &  &MP&\multicolumn{2}{c|}{GMP}& MP & GMP & ETGP\cite{F2_III,O2,B2} \\\hline
B$_2$ & 1.590 & 67.73 & 1.88 & 2.22 & 5.13 & 2.84 & 1.43 & 0.023\\
O$_2$ & 1.208 & 120.14 & 2.66 & 3.20 & 7.45 & 5.13 & 2.82 & 0.053 \\
F$_2$ & 1.411 & 38.14 & 2.85 &3.62 & 8.58 & 2.00 & 0.62 & 0.036\\\hline
\end{tabular}
\label{tab:paramsUnits}
\end{table}
\begin{table}[tbph]
\centering
\caption{Fitted parameters for the MP and GMP in reduced coordinates for B$_2$, O$_2$, and F$_2$.  The RMSD of the various potential fits and the \textit{ab-initio} PESs\cite{F2_II,O2,B2} are also indicated.\\}
\begin{tabular}{|c|c| c c|c c c|}
\hline
Molecule & $\alpha$&$\alpha$ & $\eta$ & \multicolumn{3}{c|}{RMSD}\\\hline
& MP & \multicolumn{2}{c|}{GMP}& MP & GMP & ETGP\cite{F2_III,O2,B2}\\\hline
B$_2$& 2.99 & 3.54 & 2.30 & 0.042 & 0.021 & 0.00034\\
O$_2$& 3.21 & 3.86 & 2.33 & 0.043 & 0.023 & 0.00044\\
F$_2$ & 4.01 & 5.10 & 2.37 & 0.052 & 0.016 & 0.00094\\ \hline
\end{tabular}
\label{tab:paramsReduced}
\end{table}

Examining the fitted parameters in reduced coordinates (Table \ref{tab:paramsReduced}) provides additional insights into how similar the fits of homologous molecules are. For the three diatomic molecules examined here, $\eta$ sits right in between 2.3 and 2.4, which is different than the range of values 0.22 - 0.28 previously reported for ion-water PESs.\cite{werhahn2014universal}  It is important to note that for 
$\eta$ > 2 there is a turnover point in the repulsive wall.  While for $\eta$ < 2.4 the turnover point is at short bond distances and high energies so solutions can still be obtained.  However, when $\eta$ > 2.45, there are significant concerns in the accuracy of the potential and the stability of the numerical methods.  Another interesting observation is the increased value of $\alpha$ in the GMP compared to the MP. It appears that for these covalently bonded systems $\alpha$ increases to correct the shoulder of the PES at larger $r^*$ whereas $\gamma$ rectifies the corresponding changes to the repulsive wall and the minimum of the PES.

\subsection{Application of the Laguerre Polynomial Method (LPM) to the vibrational energy levels}

The availability of highly accurate, FCI quality PESs for the B$_2$, O$_2$, and F$_2$ diatomic molecules provides the opportunity to benchmark various numerical techniques used to obtain the vibrational levels. In addition, the availability of experimentally measured vibrational levels up to dissociation provides a critical validation test for the accuracy of the computed PES. As stated previously, the vibrational levels of the MP are analytically available.  The GMP can be solved through a variety of methods (\textit{vide supra}), but this section will focus on the Laguerre Polynomial Method (LPM) outlined in section II and the Colbert Miller Discrete Variable Representation (CM-DVR). \cite{CMDVR} This section also analyses the vibrational levels obtained via the ETGP and a cubic interpolated surface (Inter) and compares the vibrational levels to the experimentally measured ones.  All calculations were performed in reduced coordinates.  The Laguerre Polynomial Method (LPM) was applied with $\sim 2\lambda$ Laguerre basis functions, since the all the vibrational states (except for the last one) were converged well within a wavenumber (Figure \ref{fig:timeConverge}).  For CM-DVR, a simulation range ($r^*$) from [0.5, 5] was initially constructed with 500 gridpoints, yielding $\Delta r^*$ = 0.009.  It should be noted that systems with large values of $\eta$, such as F$_2$, have a turnover point at $r^*$= 0.6, which is within the DVR range of [0.5, 5] used to compute the vibrational points.  To prevent unphysical eigenstates resulting from from the turnover, the potential was held to the value of the turnover to grid points less than the turnover, as shown in Figure \ref{fig:Pots} for F$_2$.

\begin{figure}[tbph]
\centering
\includegraphics[width=\textwidth]{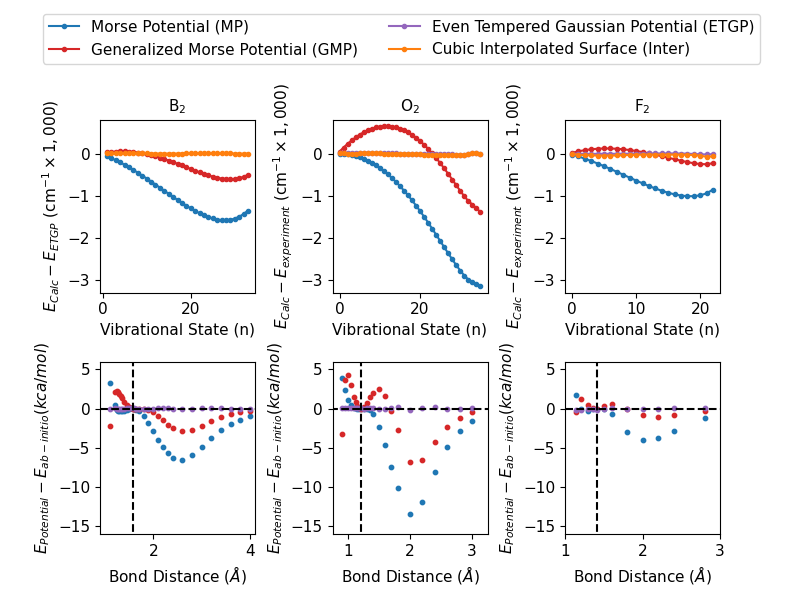}
\caption{\textbf{(Top)} Errors ($E_{Calc}-E_{Exp}$ or $E_{Calc}-E_{ETGP}$) in vibrational energies (cm$^{-1}$) plotted against the vibrational level. Each line represents the various fits of the \textit{ab-initio} PES of Bytautas \textit{et al.}\cite{F2_II,F2_III,O2,B2} (MP: Morse Potential, GMP: Generalized Morse Potentail, ETGP: Even-Tempered Gaussian Potential, Inter: cubic Interpolated Surface). \textbf{(Bottom)} Errors ($E_{ab-initio}-E_{Potential}$) (kcal/mol) in the \textit{ab-initio} points\cite{F2_II,O2,B2}.  The vertical line denotes the equilibrium geometry ($r_m$, minimum of the potential).}
\label{fig:All}
\end{figure}

The experimentally measured vibrational levels of O$_2$ and F$_2$ have been previously reported up to dissociation.  The experimental vibrational levels constitute the best observable that describes the ``experimental'' PES and serve as a powerful benchmark of the accuracy of the PES.  Previous reports have already shown the FCI \textit{ab-inito} PESs reported by Bytautas \textit{et al.} are extremely accurate in reproducing the experimental vibrations.\cite{F2_III,O2}  The simpler Laguerre based methods are more easily solvable so confirming their accuracy to experiment is advantageous. The vibrational levels of F$_2$ are reported in Table \ref{tab:F2} and their errors from experiment are displayed in Figure \ref{fig:All}.  Importantly, one achieves the same answer for the GMP using the LPM as with CM-DVR until close to dissociation where tight convergence is not achieved under these conditions. This validates the LPM for these covalent systems and the choice of using $\sim 2\lambda$ basis functions.  In addition, one notices an improvement in the higher vibrational levels and root mean square error in the GMP compared to the MP, associated with the improvement in the fit to highly accurate \textit{ab-initio} points.  Naturally, none of the schemes is as accurate as the highly parameterized ETGP used by Bytautas \textit{et al.} Nevertheless, the GMP corresponds to a good and easily solvable potential for F$_2$.

\begin{table}[tbph]
\centering
\caption{The vibrational levels of F$_2$ (cm$^{-1}$) for various potential fits to the \textit{ab-initio} PES of Bytautas \textit{et al}.\cite{F2_II} (MP: Morse Potential, GMP: Generalized Morse Potential, ETGP: Even-Tempered Gaussian Potential, Inter: cubic Interpolated Surface). Subscripts denote the method of solving for the vibrational levels (Anal: analytical, LPM: Laguerre Polynomial Method, DVR: Colbert-Miller Discrete Variable Representation). Listed Root Mean Square Errors (RMSE) and Maximum Absolute Errors (maxAE) are computed with respect to the measured experimental vibrations for F$_2$.\cite{F2_Exp}\\}
\scalebox{0.6}{
\begin{tabular}{|c|c c c c c|c|c|c|c c c c c|c|}
\hline 
$\nu$ & MP$_\text{Anal}$ & GMP$_\text{LPM}$ & GMP$_\text{DVR}$ & ETGP$_\text{DVR}$ & Inter$_\text{DVR}$ & Experiment&~& $\nu$ & MP$_\text{Anal}$ & GMP$_\text{LPM}$ & GMP$_\text{DVR}$ & ETGP$_\text{DVR}$ & Inter$_\text{DVR}$ & Experiment\\\hline
0 & 434.60 & 475.66 & 475.66 & 454.26 & 444.83 & 455.37 & ~ & 15 & 10126.17 & 10958.67 & 10958.67 & 11060.71 & 11033.24 & 11054.99 \\ 
1 & 1282.20 & 1405.17 & 1405.17 & 1346.34 & 1323.27 & 1349.27 & ~ & 16 & 10541.99 & 11376.79 & 11376.79 & 11514.07 & 11486.9 & 11509.27 \\ 
2 & 2101.02 & 2303.27 & 2303.27 & 2215.45 & 2184.64 & 2219.52 & ~ & 17 & 10929.02 & 11756.81 & 11756.81 & 11926.96 & 11900.66 & 11924.33 \\ 
3 & 2891.06 & 3169.65 & 3169.65 & 3060.96 & 3024.71 & 3065.59 & ~ & 18 & 11287.27 & 12098.03 & 12098.03 & 12296.81 & 12271.29 & 12297.99 \\ 
4 & 3652.30 & 4004.01 & 4004.01 & 3882.17 & 3841.92 & 3886.9 & ~ & 19 & 11616.73 & 12399.76 & 12399.76 & 12620.57 & 12593.82 & 12627.62 \\ 
5 & 4384.76 & 4806.00 & 4806.00 & 4678.34 & 4635.87 & 4682.80 & ~ & 20 & 11917.40 & 12661.21 & 12661.21 & 12894.52 & 12859.76 & 12908.35 \\ 
6 & 5088.44 & 5575.27 & 5575.27 & 5448.72 & 5406.63 & 5452.56 & ~ & 21 & 12189.29 & 12881.60 & 12881.60 & 13113.98 & 13059.38 & 13133.37 \\ 
7 & 5763.33 & 6311.47 & 6311.47 & 6192.47 & 6153.21 & 6195.42 & ~ & 22 & 12432.40 & 13060.07 & 13060.07 & 13272.71 & 13220.98 & 13285.75 \\ 
8 & 6409.44 & 7014.19 & 7014.19 & 6908.71 & 6873.85 & 6910.54 & ~ & 23 & 12646.72 & 13195.66 & 13195.71 & ~ & ~ & ~ \\ 
9 & 7026.76 & 7683.04 & 7683.04 & 7596.51 & 7566.11 & 7597.00 & ~ & 24 & 12832.25 & 13286.88 & 13287.55 & ~ & ~ & ~ \\ 
10 & 7615.29 & 8317.57 & 8317.57 & 8254.81 & 8227.43 & 8253.85 & ~ & 25 & 12988.99 & 13329.82 & 13334.59 & ~ & ~ & ~ \\ 
11 & 8175.04 & 8917.34 & 8917.34 & 8882.51 & 8856.33 & 8880.04 & ~ & 26 & 13116.96 & ~ & ~ & ~ & ~ & ~ \\ 
12 & 8706.00 & 9481.86 & 9481.86 & 9478.37 & 9452.30 & 9474.48 & ~ & 27 & 13216.13 & ~ & ~ & ~ & ~ & ~ \\ 
13 & 9208.17 & 10010.62 & 10010.62 & 10041.06 & 10014.47 & 10036.00 & ~ & 28 & 13286.52 & ~ & ~ & ~ & ~ & ~ \\ 
14 & 9681.56 & 10503.08 & 10503.08 & 10569.07 & 10541.87 & 10563.39 & ~ & ~ & ~ & ~ & ~ & ~ & ~ & ~ \\ \hline
RMSD & 706.52 & 136.56 & 136.56 & 6.75 & 37.41 & & & maxAE & 1010.89 & 251.77 & 251.77 & 19.39 & 17.99 &  \\\hline
\end{tabular}}
\label{tab:F2}
\end{table}

In a similar manner, the vibrational levels of O$_2$ were calculated and compared to experiment. The vibrational levels are reported in Table \ref{tab:O2} and the errors are shown in Figure \ref{fig:All}.  Again, the vibrational levels obtained with the GMP are the same through the LPM and CM-DVR methods up to the last three vibrational levels, further supporting the validity of the LPM.  Although the GMP fits the \textit{ab-initio} PES of O$_2$ better than the MP, it does not accurately predict the the vibrational levels, with RMSEs well above 500 cm$^{-1}$.  The errors in the the GMP vibrational levels originate from the poor description of the dissociative shoulder and the minimum. The MP seems to predict better lower energy vibrational levels than the GMP. It appears that the large errors in the MP's long-range regime cause the GMP to sacrifice a lot in the minimum area to fix more in the shoulder (Figure \ref{fig:All}, bottom). This is purely a result of the individual PES, since it is much less Morse-like.

\begin{table}[tbph]
\centering
\caption{The vibrational levels of O$_2$ (cm$^{-1}$) for various potential fits to the \textit{ab-initio} PES of Bytautas \textit{et al}.\cite{O2}. The definitions of the various symbols and acronyms are the same as in Table \ref{tab:F2}. Listed Root Mean Square Errors (RMSE) and Maximum Absolute Errors (maxAE) are computed with respect to the measured experimental vibrations for O$_2$.\cite{O2_Exp_0-22,O2_Exp_0-28,O2_Exp_26-31,O2_Exp_29-35}\\}
\scalebox{0.6}{
\begin{tabular}{|c|c c c c c|c|c|c|c c c c c |c|}
\hline 
$\nu$ & MP$_\text{Anal}$ & GMP$_\text{LPM}$ & GMP$_\text{DVR}$ & ETGP$_\text{DVR}$ & Inter$_\text{DVR}$ & Experiment&&$\nu$ & MP$_\text{Anal}$ & GMP$_\text{LPM}$ & GMP$_\text{DVR}$ & GP$_\text{DVR}$ & Inter$_\text{DVR}$ & Experiment \\\hline
0 & 787.47 & 836.94 & 836.94 & 791.64 & 791.45 & 787.20 & ~ & 26 & 31472.04 & 33345.03 & 33345.03 & 33683.36 & 33652.65 & 33695.57 \\ 
1 & 2340.07 & 2488.88 & 2488.88 & 2355.55 & 2347.80 & 2343.59 & ~ & 27 & 32250.00 & 34138.20 & 34138.20 & 34605.32 & 34577.00 & 34620.31 \\ 
2 & 3862.88 & 4110.08 & 4110.08 & 3894.05 & 3877.33 & 3876.31 & ~ & 28 & 32998.16 & 34895.34 & 34895.34 & 35494.06 & 35470.17 & 35510.89 \\ 
3 & 5355.89 & 5700.39 & 5700.39 & 5407.46 & 5387.90 & 5385.81 & ~ & 29 & 33716.53 & 35616.17 & 35616.17 & 36347.45 & 36329.80 & 36369.13 \\ 
4 & 6819.11 & 7259.65 & 7259.65 & 6896.11 & 6875.30 & 6873.89 & ~ & 30 & 34405.11 & 36300.41 & 36300.41 & 37162.99 & 37152.68 & 37186.83 \\ 
5 & 8252.53 & 8787.70 & 8787.70 & 8360.33 & 8339.59 & 8334.89 & ~ & 31 & 35063.89 & 36947.76 & 36947.76 & 37937.66 & 37934.45 & 37962.58 \\ 
6 & 9656.16 & 10284.38 & 10284.38 & 9800.47 & 9781.02 & 9775.34 & ~ & 32 & 35692.88 & 37557.94 & 37557.94 & 38667.75 & 38669.64 & 38684.20 \\ 
7 & 11030.00 & 11749.52 & 11749.52 & 11216.85 & 11199.59 & 11194.08 & ~ & 33 & 36292.07 & 38130.63 & 38130.63 & 39348.62 & 39352.30 & 39339.20 \\ 
8 & 12374.04 & 13182.95 & 13182.95 & 12609.75 & 12595.29 & 12588.87 & ~ & 34 & 36861.47 & 38665.53 & 38665.53 & 39974.23 & 39975.71 & 39967.20 \\ 
9 & 13688.29 & 14584.50 & 14584.50 & 13979.46 & 13967.90 & 13960.80 & ~ & 35 & 37401.08 & 39162.30 & 39162.30 & 40536.37 & 40531.09 & 40547.20 \\ 
10 & 14972.74 & 15953.99 & 15953.99 & 15326.20 & 15317.17 & 15310.44 & ~ & 36 & 37910.89 & 39620.64 & 39620.64 & 41023.12 & 41007.05 & ~ \\ 
11 & 16227.40 & 17291.24 & 17291.24 & 16650.15 & 16643.02 & 16639.62 & ~ & 37 & 38390.91 & 40040.18 & 40040.18 & 41415.58 & 41386.76 & ~ \\ 
12 & 17452.27 & 18596.07 & 18596.07 & 17951.43 & 17945.62 & 17943.76 & ~ & 38 & 38841.14 & 40420.58 & 40420.58 & 41682.68 & 41655.58 & ~ \\ 
13 & 18647.34 & 19868.29 & 19868.29 & 19230.12 & 19225.18 & 19224.24 & ~ & 39 & 39261.57 & 40761.49 & 40761.49 & 41822.31 & 41828.49 & ~ \\ 
14 & 19812.62 & 21107.70 & 21107.70 & 20486.21 & 20481.63 & 20482.40 & ~ & 40 & 39652.20 & 41062.53 & 41062.53 & 41921.19 & 41966.69 & ~ \\ 
15 & 20948.10 & 22314.10 & 22314.10 & 21719.63 & 21714.89 & 21718.10 & ~ & 41 & 40013.05 & 41323.32 & 41323.32 & 41996.26 & ~ & ~ \\ 
16 & 22053.80 & 23487.30 & 23487.30 & 22930.22 & 22924.77 & 22930.30 & ~ & 42 & 40344.10 & 41543.48 & 41543.48 & ~ & ~ & ~ \\ 
17 & 23129.69 & 24627.09 & 24627.09 & 24117.78 & 24110.97 & 24118.70 & ~ & 43 & 40645.35 & 41722.59 & 41722.59 & ~ & ~ & ~ \\ 
18 & 24175.80 & 25733.25 & 25733.25 & 25281.97 & 25273.09 & 25284.31 & ~ & 44 & 40916.81 & 41860.23 & 41860.24 & ~ & ~ & ~ \\ 
19 & 25192.10 & 26805.57 & 26805.57 & 26422.40 & 26410.66 & 26425.73 & ~ & 45 & 41158.48 & 41955.86 & 41956.01 & ~ & ~ & ~ \\ 
20 & 26178.62 & 27843.84 & 27843.84 & 27538.57 & 27523.22 & 27542.66 & ~ & 46 & 41370.35 & 42007.80 & 42009.46 & ~ & ~ & ~ \\ 
21 & 27135.34 & 28847.81 & 28847.81 & 28629.87 & 28610.42 & 28634.70 & ~ & 47 & 41552.43 & ~ & ~ & ~ & ~ & ~ \\ 
22 & 28062.27 & 29817.26 & 29817.26 & 29695.57 & 29672.03 & 29701.25 & ~ & 48 & 41704.72 & ~ & ~ & ~ & ~ & ~ \\ 
23 & 28959.40 & 30751.95 & 30751.95 & 30734.84 & 30707.68 & 30741.78 & ~ & 49 & 41827.21 & ~ & ~ & ~ & ~ & ~ \\ 
24 & 29826.74 & 31651.65 & 31651.65 & 31746.69 & 31716.84 & 31754.78 & ~ & 50 & 41919.91 & ~ & ~ & ~ & ~ & ~ \\ 
25 & 30664.29 & 32516.09 & 32516.09 & 32729.97 & 32698.82 & 32739.82 & ~ & 51 & 41982.81 & ~ & ~ & ~ & ~ & ~ \\ \hline
RMSE & 1686.51 & 642.24 & 642.24 & 17.66 & 21.37 & ~& ~ & MaxAE & 3146.12 & 1384.90 & 1384.90 & 34.98 & 43.31 & ~ \\ \hline
\end{tabular}}
\label{tab:O2}
\end{table}

Finally, the vibrational levels of $^{11}$B$_2$ were calculated and compared to those reported by Bytautas \textit{et al.}\cite{B2}.  There is no full vibrational analysis of B$_2$ and only certain energies have been reported.\cite{CanadianBoron,OldBoron,NewBoron} Since the comparison of the ETGP to the available experimental values has already been reported\cite{B2}, we will focus solely on comparing the GMP and the MP to the ETGP.  The ETGP was chosen as the reference for this system since it has been shown to be highly accurate for O$_2$ and F$_2$. The vibrational levels for $^{11}$B$_2$ are reported in Table \ref{tab:B2}.  It must be pointed out again that the same vibrational energies are obtained via the LPM and CM-DVR methods. Since the full experimental vibrational levels for $^{11}$B$_{2}$ are not available, we will investigate the accuracy of the MP and the GMP by comparing the results with these potentials to the ones previously reported by Bytautas \textit{et al.}\cite{B2} (Figure \ref{fig:All}).  The differences for B$_2$ are much smaller for GMP than MP for all vibrational states in ways similar to F$_2$.  The GMP, again, is a much better fit to the \textit{ab-initio} surfaces than the MP and is an excellent fit for F$_2$ and B$_2$.

\begin{table}[tbph]
\centering
\caption{The vibrational levels of $\ ^{11}$B$_2$ (cm$^{-1}$) for various potential fits to the \textit{ab-initio} PES of Bytautas \textit{et al}.\cite{B2}. The definitions of the various symbols and acronyms are the same as in Table \ref{tab:F2}.\\}
\scalebox{0.8}{
\begin{tabular}{|c|c c c c c|c|c|c c c c c|}
\hline ~~~$\nu$~~~ & MP$_\text{Anal}$ & GMP$_\text{LPM}$ & GMP$_\text{DVR}$ & 
ETGP$_\text{DVR}$ & Inter$_\text{DVR}$ & ~ & ~~~$\nu$~~~ & MP$_\text{Anal}$ & GMP$_\text{LPM}$ & GMP$_\text{DVR}$ & ETGP$_\text{DVR}$ & Inter$_\text{DVR}$\\\hline
0 & 503.36 & 533.19 & 533.19 & 522.52 & 523.72 & ~ & 23 & 17815.12 & 18798.73 & 18798.73 & 19279.01 & 19288.76 \\ 
1 & 1493.87 & 1583.46 & 1583.46 & 1554.44 & 1558.50 & ~ & 24 & 18308.37 & 19299.88 & 19299.88 & 19814.62 & 19825.15 \\ 
2 & 2462.75 & 2611.36 & 2611.36 & 2568.87 & 2574.84 & ~ & 25 & 18779.99 & 19775.20 & 19775.20 & 20320.62 & 20331.37 \\ 
3 & 3410.02 & 3616.78 & 3616.78 & 3568.87 & 3573.12 & ~ & 26 & 19230.00 & 20224.49 & 20224.49 & 20795.73 & 20806.02 \\ 
4 & 4335.66 & 4599.59 & 4599.59 & 4544.39 & 4552.73 & ~ & 27 & 19658.38 & 20647.55 & 20647.55 & 21238.58 & 21247.73 \\ 
5 & 5239.69 & 5559.68 & 5559.68 & 5505.03 & 5513.54 & ~ & 28 & 20065.15 & 21044.17 & 21044.17 & 21647.63 & 21655.11 \\ 
6 & 6122.09 & 6496.91 & 6496.91 & 6447.25 & 6455.54 & ~ & 29 & 20450.29 & 21414.13 & 21414.13 & 22021.22 & 22026.64 \\ 
7 & 6982.88 & 7411.16 & 7411.16 & 7370.77 & 7378.41 & ~ & 30 & 20813.82 & 21757.22 & 21757.22 & 22357.59 & 22360.73 \\ 
8 & 7822.04 & 8302.31 & 8302.31 & 8275.30 & 8282.05 & ~ & 31 & 21155.72 & 22073.20 & 22073.20 & 22654.98 & 22656.17 \\ 
9 & 8639.59 & 9170.21 & 9170.21 & 9160.30 & 9166.27 & ~ & 32 & 21476.01 & 22361.84 & 22361.84 & 22911.89 & 22912.14 \\ 
10 & 9435.51 & 10014.74 & 10014.74 & 10026.09 & 10030.75 & ~ & 33 & 21774.68 & 22622.89 & 22622.89 & 23127.55 & 23128.70 \\ 
11 & 10209.82 & 10835.75 & 10835.75 & 10871.65 & 10875.29 & ~ & 34 & 22051.72 & 22856.12 & 22856.12 & 23302.78 & 23306.47 \\ 
12 & 10962.50 & 11633.11 & 11633.11 & 11696.79 & 11699.60 & ~ & 35 & 22307.15 & 23061.27 & 23061.27 & 23440.77 & 23447.01 \\ 
13 & 11693.57 & 12406.66 & 12406.66 & 12501.11 & 12503.27 & ~ & 36 & 22540.95 & 23238.06 & 23238.06 & 23546.61 & 23553.17 \\ 
14 & 12403.02 & 13156.26 & 13156.26 & 13284.15 & 13285.95 & ~ & 37 & 22753.14 & 23386.23 & 23386.23 & 23624.78 & 23630.13 \\ 
15 & 13090.84 & 13881.76 & 13881.76 & 14045.42 & 14047.15 & ~ & 38 & 22943.70 & 23505.48 & 23505.49 & 23676.03 & 23674.85 \\ 
16 & 13757.05 & 14583.01 & 14583.01 & 14784.40 & 14786.39 & ~ & 39 & 23112.65 & 23595.53 & 23595.55 & ~ & ~ \\ 
17 & 14401.63 & 15259.84 & 15259.84 & 15500.53 & 15503.14 & ~ & 40 & 23259.97 & 23655.76 & 23656.11 & ~ & ~ \\ 
18 & 15024.60 & 15912.09 & 15912.09 & 16193.18 & 16196.77 & ~ & 41 & 23385.68 & 23683.95 & 23686.90 & ~ & ~ \\ 
19 & 15625.94 & 16539.60 & 16539.60 & 16861.70 & 16866.52 & ~ & 42 & 23489.76 & ~ & ~ & ~ & ~ \\ 
20 & 16205.67 & 17142.19 & 17142.19 & 17505.37 & 17511.52 & ~ & 43 & 23572.23 & ~ & ~ & ~ & ~ \\ 
21 & 16763.77 & 17719.70 & 17719.70 & 18123.39 & 18130.84 & ~ & 44 & 23633.07 & ~ & ~ & ~ & ~ \\ 
22 & 17300.26 & 18271.94 & 18271.94 & 18714.92 & 18723.58 & ~ & ~ & ~ & ~ & ~ & ~ & ~ \\ \hline
\end{tabular}}
\label{tab:B2}
\end{table}

The trends in the errors in the vibrational levels for all three diatomic molecules, obtained with the MP, GMP, ETGP and interpolated methods, can be analyzed with the visual help of Figure \ref{fig:All}.  The MP underestimates the vibrational levels for all states to various extents arising from large understimations of the shoulder immediately after the equilibrium geometry.  The GMP overestimates the lower energy states and underestimates the higher energy states.  The additional exponential in the GMP sacrifices accuracy at the minimum to improve the overall fit and the vibrational levels near dissociation, as seen in the lower plots of Figure \ref{fig:All}. This effect is further exaggerated for less Morse-like potentials, such as O$_2$, where the large error near dissociaton is greatly reduced at the cost of introducing noticeable errors near the minimum.  The ETGP does extremely well predicting all vibrational states and all \textit{ab-initio} points because it is highly parameterized.  Similarly, the Inter surface does well for B$_2$ and O$_2$, but not for F$_2$, which further highlights the important considerations of sampling: there are only 13 \textit{ab-initio} points for F$_2$, a fact that results in over-fitting of the Inter surface.  The errors in the vibrational levels can be directly attributed to the errors in the PES and this is an important consideration when solving for the vibrational levels of improved potential energy functions.

\section{Synopsis}
We have presented several methods for solving the vibrational problem of the GMP, which represents a minimal extension the original MP amounting to an additional exponential term and a constant at a cost of sacrificing the existence of an analytical solution for the vibrational energy levels. For this reason, we have introduced Galerkin schemes such as the Laguerre Polynomial Method (LPM), the Symmetrized Laguerre Polynomial Method (SLPM), and the Polynomial Expansion method (PEM) to solve for the vibrational levels of the GMP, and compared the results and their timing with the ones for the original MP and th CM-DVR method. The above methods were applied to the diatomic molecules B$_2$, O$_2$ and F$_2$, for which theoretically computed FCI-quality PESs and experimentally measured vibrational levels up to dissociation have been reported. There is a clear improvement of the fit of the ab-initio points with the GMP over the MP, resulting in an overall 50\% reduction in the error. 

We found that the LPM produces vibrational states for the GMP that are converged to within spectroscopic accuracy (0.01 cm$^{-1}$) at 1 -- 2 orders of magnitude faster while requiring much fewer degrees of freedom than the Colbert-Miller Discrete Variable Representation (CM-DVR) method. It is important to note that for the GMP, the same answer is reached using the LPM compared to the more expensive CM-DVR method for all vibrational levels up to dissociation. Since vibrational states close to dissociation are less Morse-like, the corresponding eigenfunctions are harder to approximate for the LPM using basis functions based on solutions of the MP.

The LPM and the SLPM are a full order of magnitude faster than the literature standard, CM-DVR. The LPM is the fastest method for finding the eigenstates for the GMP. The SLPM is comparable to the LPM, but suffers from numerical stability issues, and thus must be used with caution. DVR can achieve the greatest accuracy because it is highly stable, but this comes at increased cost. It is important, however, to recognize that CM-DVR is transferable to any PES while LPM and SLPM are specific to Morse-like potentials.

The underestimation of the vibrational levels with the MP is due to the poor description of the “shoulder” immediately after the equilibrium distance. On the other hand, the GMP was found to overestimate the lower vibrational levels while underestimating the higher ones. This is due to the “trade-off” between describing the area around the minimum and the rest of the PES, especially the “shoulder” close to dissociation. Nevertheless, this simple extension of the MP with just one additional parameter represents a significant improvement while maintaining a physically intuitive picture of a simple universal function that can be used to accurately describe complex PESs.

\section{Acknowledgments}
GDS and SSX were supported from the Center for Scalable Predictive methods for Excitations and Correlated phenomena (SPEC), which is funded by the U.S. Department of Energy, Office of Science, Basic Energy Sciences, Chemical Sciences, Geosciences and Biosciences Division as part of the Computational Chemical Sciences (CCS) program at Pacific Northwest National Laboratory. The work of SQ is supported by the Department of Energy (DOE) Office of Advanced Scientific Computing Research (ASCR) through the ASCR Distinguished Computational Mathematics Postdoctoral Fellowship. Battelle operates the Pacific Northwest National Laboratory for the U.S. Department of Energy. This research used computer resources provided by the National Energy Research Scientific Computing Center, which is supported by the Office of Science of the U.S. Department of Energy under Contract No. DE-AC02-05CH11231.

\vspace{1.0cm}

\noindent
\textbf{AVAILABILITY OF DATA}

The data that support the findings of this study are available from the corresponding author upon reasonable request.

\appendix 

\section{The Symmetrized Laguerre Polynomial Method (SLPM)}\label{SecSymm}
A possible pitfall of using the technique outlined in Section \ref{SecLagExp} is that the matrix (\ref{SMat}) that we end up with will always be non-symmetric. As a result, the eigenvalues produced by the algorithm are not guaranteed to be real, especially when the parameters are pushed to extreme cases. To circumvent this, we make use of a different inner product that exploits the structure of (\ref{ScaGM2}) to always yield a symmetric matrix. We define
\eqn{
\ip{f_1,f_2}_D = \int_0^{\infty} \overline{f_1(z)} f_2(z) \frac{1}{z} \ dz, \label{IPD}
}

\noindent rewrite (\ref{ScaGM2}) as
\eqn{
z\pz\left(z \pz \tilde{\psi}(z)\right) + q_{\eta}(z)\tilde{\psi}(z) = -\epsilon \tilde{\psi}(z), \label{ScaGM3}
}
and impose the Galerkin condition
\eqn{
\ip{u_l^{(\beta)} , z\pz\left(z \pz \tilde{\psi}\right) + q_{\eta}(z)\tilde{\psi}}_D =  -\epsilon \ip{u_l^{(\beta)} , \tilde{\psi}}_D, \label{GalS}
}
for $l \geq 0$ with respect to the new inner product, where $\{u_l^{(\beta)}(z)\}_{l \geq 0}$ are the basis functions defined as in Section \ref{SecLagExp}, by $u_l^{(\beta)}(z) = e^{-z/2}z^{\beta/2}p_l(z)$ for some  family of polynomials $\{p_l\}_{l \geq 0}$. Noting that
\eqn{
&& \ip{u_l^{(\beta)} , z\pz\left(z \pz \tilde{\psi}\right)}_D \nonumber\\
&=& \int_0^{\infty} u_l^{(\beta)}(z) z\pz \left(z \pz \tilde{\psi}(z)\right) \frac{1}{z} \ dz \nonumber\\
&=& \int_0^{\infty} u_l^{(\beta)}(z) \pz \left(z \pz \tilde{\psi}(z)\right) \ dz \nonumber\\
&=& \underbrace{\left[zu_l^{(\beta)}(z)  \pz \tilde{\psi}(z)\right]_0^{\infty}}_{\text{zero as $\lim_{z \to \infty} u_l^{(\beta)}(z) = 0$}}-\int_0^{\infty} z \pz u_l^{(\beta)}(z)  \pz \tilde{\psi}(z) \ dz, \label{GalS1}
}
we end up with
\eqn{
 \int_0^{\infty} -z \pz u_l^{(\beta)}(z)  \pz \tilde{\psi}(z) + \frac{q_{\eta}(z)}{z} u_l^{(\beta)}(z)  \tilde{\psi}(z) \ dz 
&=& -\epsilon \int_0^{\infty} u_l^{(\beta)}(z)  \tilde{\psi}(z) \frac{1}{z}\ dz. \label{GalS2}
}

Next, plugging the expansion $\tilde{\psi}(z) = \sum_{j = 0}^M c_j u_j^{(\beta)}(z)$ and using only $0 \leq l \leq M$ leads to
\eqn{
\sum_{j = 0}^M c_j \int_0^{\infty} -z \pz u_l^{(\beta)}(z)  \pz u_j^{(\beta)}(z)  + \frac{q_{\eta}(z)}{z} u_l^{(\beta)}(z)  u_j^{(\beta)}(z)  \ dz 
&=& -\epsilon \sum_{j = 0}^M c_j \int_0^{\infty} u_l^{(\beta)}(z)  u_j^{(\beta)}(z) \frac{1}{z}\ dz. \label{GalS3}
}

This can be recognized as the generalized eigenvalue problem
\eqn{
R^{(\eta)}\tb{c} = -\epsilon T\tb{c}, \label{GalS4}
}
for the coefficients $\tb{c} = (c_j)$ and the $(M+1) \times (M+1)$ matrices $R^{(\eta)}$ and $T$ defined by
\eqn{
&& R^{(\eta)}_{lj} = \int_0^{\infty} -z \pz u_l^{(\beta)}(z)  \pz u_j^{(\beta)}(z)  + \frac{q_{\eta}(z)}{z} u_l^{(\beta)}(z)  u_j^{(\beta)}(z)  \ dz, \nonumber\\
&& T_{lj} = \int_0^{\infty} u_l^{(\beta)}(z)  u_j^{(\beta)}(z) \frac{1}{z}\ dz, \label{RTdefnS}
}
where $0 \leq l,j \leq M$. Observe that in order for these integrals to converge, we require $\beta \geq 1$. Further note that
\eqn{
T_{lj} = \int_0^{\infty} e^{-z}z^{\beta-1} p_l(z) p_j(z) \ dz, \label{TsimpS}
}
so choosing $\{p_l\}$ to be the orthonormal generalized Laguerre polynomials with Laguerre parameter $(\beta-1)$ leads to to $T = I$, the identity matrix. Thus, (\ref{GalS4}) reduces to a symmetric eigenvalue problem (i.e., not to a generalized eigenvalue problem).

\section{The Polynomial Expansion Method (PEM) in a Finite Domain}\label{SecFinite}
One potential weakness shared by the two methods based on Laguerre expansions is that they presuppose the rate of decay of the eigenfunctions for large $z$. In cases where this decay occurs at rates that are significantly different from $e^{-z/2}$, those approaches may fail. To address this issue, we make use of the observation that, in fact, the largest domain (\ref{ScaSchrod}) needs to be solved on is $(0,\infty)$. Under the transformation $z = 2\lambda e^{-(x-x_m)}$, the corresponding domain for (\ref{ScaGM2}) is $(0,2\lambda e^{x_m})$. Thus, strictly speaking, the eigenvalue problem only needs to be solved on a finite domain, in contrast with the exact solution in the classical case $\eta = 1$ or the approaches explored in the earlier sections, all of which yield the solutions of (\ref{ScaGM2}) on $(0,\infty)$.     

This insight is helpful because, on a finite domain, supplementing the eigenvalue problem with boundary conditions allows us to solve for the unknown eigenfunctions in terms of polynomials without appending any additional decay factors. The suitable boundary conditions can be deduced by considering the corresponding end-points for (\ref{ScaSchrod}): $z = 0$ corresponds to $x = \infty$, at which point the wavefunction must vanish. The other end is more involved: as our non-dimensionalization does not get rid of $x_m$ and we do not have a handle on its value, we make do by replacing the right end-point $z = 2\lambda e^{x_m}$ by $z = z_0$ for a large enough $z_0$; furthermore, taking inspiration from the rapid decay of the solutions, we impose a homogeneous Dirichlet condition at this end-point too. To sum up, we need to solve
\eqn{
\left\{
\begin{matrix}
z \pz \left(z \pz \tilde{\psi}(z)\right) + q_{\eta}(z)\tilde{\psi}(z) = -\epsilon \tilde{\psi}(z), & z \in (0,z_0), \\
\tilde{\psi}(0) = \tilde{\psi}(z_0) = 0. 
\end{matrix} \right. \label{FiniteEig}
}

The numerical procedure is fairly similar to the ones employed earlier. We choose an appropriate set of basis functions $\{r_j(z)\}_{j \in \mathcal{J}}$ on $[0,z_0]$, expand the eigenfunctions $\tilde{\psi}(z) = \sum_{j \in \mathcal{J}}c_j r_j(z)$, and apply the Galerkin condition
\eqn{
\ip{r_l,z \pz \left(z \pz \tilde{\psi}\right) + q_{\eta}(z)\tilde{\psi}} = -\epsilon \ip{r_l,\tilde{\psi}}, \label{GalCond2}
}
for $l \in \mathcal{J}$, where $\ip{\cdot,\cdot}$ is an appropriate inner product.  

Building on the symmetric procedure employed in Section \ref{SecSymm}, we again use an inner product with the $1/z$ weight function, to wit,
\eqn{
\ip{f_1,f_2}_{D,z_0} = \int_0^{z_0} \overline{f_1(z)} f_2(z) \frac{1}{z} \ dz. \label{IPDz0}
}

However, this choice could lead to singular integrals because the $1/z$ weight blows up at 0; fortunately, the boundary conditions enable us to nip this issue in the bud. Let $\{q_j\}_{j\geq 0}$ be the orthonormal Legendre polynomials on $[0,z_0]$, and define
\eqn{
r_j(z) = \left\{ \begin{matrix}
q_j(z) - \delta_j q_0(z), & \text{for even } j, \\
q_j(z) - \delta_j q_1(z), & \text{for odd } j, \\
\end{matrix} \right. \label{ClampedLegs}
} 
where 
\eqn{
\delta_j = \left\{ \begin{matrix}
\sqrt{2j+1}, & \text{for even } j, \\
\sqrt{(2j+1)/3}, & \text{for odd } j, \\
\end{matrix} \right. \label{DeltaVals}
} 
have been chosen so that $r_j(0) = r_j(z_0) = 0$ for all $j$. Note that $r_0 \equiv r_1 \equiv 0$ so our basis function indices in fact begin at $j = 2$. Applying (\ref{GalCond2}) with the inner product (\ref{IPDz0}) then gives
\eqn{
\int_0^{z_0} r_l(z) \left(z \pz \left(z \pz \tilde{\psi}(z)\right) + q_{\eta}(z)\tilde{\psi}(z)\right) \frac{1}{z} \ dz 
&=& -\epsilon \int_0^{z_0} r_l(z) \tilde{\psi}(z) \frac{1}{z} \ dz. \label{GalCond2a}
}

Simplifying yields
\eqn{
\left[zr_l(z)\pz \tilde{\psi}(z)\right]_0^{z_0} + \int_0^{z_0} -z r_l'(z) \pz \tilde{\psi}(z) + \frac{q_{\eta}(z)}{z}r_l(z) \tilde{\psi}(z) \ dz 
&=& -\epsilon \int_0^{z_0} r_l(z) \tilde{\psi}(z) \frac{1}{z} \ dz. \label{GalCond2b}
}

Using $r_l(0) = r_l(z_0) = 0$, we obtain
\eqn{
\sum_{j = 2}^M c_j \int_0^{z_0} -z r_l'(z) r_j'(z) + \frac{q_{\eta}(z)}{z}r_l(z) r_j(z) \ dz 
&=& -\epsilon \sum_{j = 2}^M c_j  \int_0^{z_0} r_l(z) r_j(z) \frac{1}{z} \ dz. \label{GalCond2c}
}

Defining the $(M-1) \times (M-1)$ matrices $R^{(\eta,z_0)}$ and $T^{(z_0)}$ by
\eqn{
&& R^{(\eta,z_0)}_{lj} = \int_0^{z_0} -z r_l'(z) r_j'(z) + \frac{q_{\eta}(z)}{z}r_l(z) r_j(z) \ dz, \nonumber\\
&& T^{(z_0)}_{lj} = \int_0^{z_0} r_l(z) r_j(z) \frac{1}{z} \ dz, \label{RTdefn}
}
for $2 \leq l,j \leq M$, and writing $\tb{c} = \left(c_j \right)$ allows us to identify (\ref{GalCond2c}) as the generalized eigenvalue problem
\eqn{
R^{(\eta,z_0)} \tb{c} = -\epsilon T^{(z_0)}\tb{c}. \label{GenEig}
}

Observe that the matrices $R^{(\eta,z_0)}$ and $T^{(z_0)}$ are symmetric and symmetric positive-definite respectively, which implies that all eigenvalues of (\ref{GenEig}) are guaranteed to be real. Moreover, as the basis functions vanish at $z = 0$, they are divisible by $z$; thus, the integrands $\left(\frac{r_l(z)r_j(z)}{z}\right)$ are polynomials and can be integrated exactly using Gaussian quadrature. This also holds for $\left(\frac{q_{\eta}(z)}{z}r_l(z) r_j(z)\right)$ when $\eta$ is an integer.

\bibliography{aipsamp2}

\begin{thebibliography}{10}
\expandafter\ifx\csname url\endcsname\relax
  \def\url#1{\texttt{#1}}\fi
\expandafter\ifx\csname urlprefix\endcsname\relax\def\urlprefix{URL }\fi
\expandafter\ifx\csname href\endcsname\relax
  \def\href#1#2{#2} \def\path#1{#1}\fi

\bibitem{morse1929diatomic}
P.~M. Morse, \href{https://link.aps.org/doi/10.1103/PhysRev.34.57}{Diatomic
  molecules according to the wave mechanics. {II}. {V}ibrational levels}, Phys.
  Rev. 34 (1929) 57--64.
\newblock \href {https://doi.org/10.1103/PhysRev.34.57}
  {\path{doi:10.1103/PhysRev.34.57}}.
\newline\urlprefix\url{https://link.aps.org/doi/10.1103/PhysRev.34.57}

\bibitem{LeRoy1}
R.~J. Le~Roy, Y.~Huang, C.~Jary, \href{https://doi.org/10.1063/1.2354502}{An
  accurate analytic potential function for ground-state {N}$_{2}$ from a
  direct-potential-fit analysis of spectroscopic data}, J. Chem. Phys. 125
  (2006) 164310.
\newblock \href {https://doi.org/10.1063/1.2354502}
  {\path{doi:10.1063/1.2354502}}.
\newline\urlprefix\url{https://doi.org/10.1063/1.2354502}

\bibitem{LeRoy2}
R.~J. Le~Roy, R.~D.~E. Henderson,
  \href{https://doi.org/10.1080/00268970701241656}{A new potential function
  form incorporating extended long-range behaviour: application to ground-state
  {C}a$_{2}$}, Mol. Phys. 105 (2007) 663.
\newblock \href {https://doi.org/10.1080/00268970701241656}
  {\path{doi:10.1080/00268970701241656}}.
\newline\urlprefix\url{https://doi.org/10.1080/00268970701241656}

\bibitem{LeRoy3}
H.~Salami, A.~J. Ross, P.~Crozet, W.~Jastrzebski, P.~Kowalczyk, R.~J. Le~Roy,
  \href{https://doi.org/10.1063/1.2734973}{A full analytic potential energy
  curve for the $\alpha^{3}{\Sigma}^{+}$ state of {KL}i from a limited
  vibrational data set}, J. Chem. Phys. 126 (2007) 194313.
\newblock \href {https://doi.org/10.1063/1.2734973}
  {\path{doi:10.1063/1.2734973}}.
\newline\urlprefix\url{https://doi.org/10.1063/1.2734973}

\bibitem{LeRoy4}
A.~Shayesteh, R.~D.~E. Henderson, R.~J. Le~Roy, P.~F. Bernath,
  \href{https://doi.org/10.1021/jp075704a}{Ground state potential energy curve
  and dissociation energy of {M}g{H}}, J. Phys. Chem. A. 111 (2007) 12495.
\newblock \href {https://doi.org/10.1021/jp075704a}
  {\path{doi:10.1021/jp075704a}}.
\newline\urlprefix\url{https://doi.org/10.1021/jp075704a}

\bibitem{LeRoy5}
R.~J. Le~Roy, N.~S. Dattani, J.~A. Coxon, A.~J. Ross, P.~Crozet, C.~Linton,
  Accurate analytic potentials for {L}i$_2$ ({X}$^1{\Sigma}_g^+$) and {L}i$_2$
  ({A}$^1{\Sigma}_u^+$) from 2 to 90 \r{A}, and the radiative lifetime of {L}i
  ($2p$), The Journal of chemical physics 131~(20) (2009) 204309.

\bibitem{xantheas2014universal}
S.~S. Xantheas, J.~C. Werhahn,
  \href{https://doi.org/10.1063/1.4891819}{Universal scaling of potential
  energy functions describing intermolecular interactions. {I}. {F}oundations
  and scalable forms of new generalized {M}ie, {L}ennard-{J}ones, {M}orse, and
  {B}uckingham exponential-6 potentials}, J. Chem. Phys. 141~(6) (2014) 064117.
\newblock \href {https://doi.org/10.1063/1.4891819}
  {\path{doi:10.1063/1.4891819}}.
\newline\urlprefix\url{https://doi.org/10.1063/1.4891819}

\bibitem{werhahn2014universal}
J.~C. Werhahn, D.~Akase, S.~S. Xantheas,
  \href{https://doi.org/10.1063/1.4891820}{Universal scaling of potential
  energy functions describing intermolecular interactions. {II}. {T}he
  halide-water and alkali metal-water interactions}, J. Chem. Phys. 141~(6)
  (2014) 064118.
\newblock \href {https://doi.org/10.1063/1.4891820}
  {\path{doi:10.1063/1.4891820}}.
\newline\urlprefix\url{https://doi.org/10.1063/1.4891820}

\bibitem{CMDVR}
D.~T. Colbert, W.~H. Miller, \href{https://doi.org/10.1063/1.462100}{A novel
  discrete variable representation for quantum mechanical reactive scattering
  via the {S}‐matrix {K}ohn method}, J. Chem. Phys. 96~(3) (1992) 1982--1991.
\newblock \href {https://doi.org/10.1063/1.462100}
  {\path{doi:10.1063/1.462100}}.
\newline\urlprefix\url{https://doi.org/10.1063/1.462100}

\bibitem{PODVR}
J.~Echave, D.~C. Clary,
  \href{https://www.sciencedirect.com/science/article/pii/000926149285330D}{Potential
  optimized discrete variable representation}, Chem. Phys. Letts. 190~(3)
  (1992) 225--230.
\newblock \href {https://doi.org/https://doi.org/10.1016/0009-2614(92)85330-D}
  {\path{doi:https://doi.org/10.1016/0009-2614(92)85330-D}}.
\newline\urlprefix\url{https://www.sciencedirect.com/science/article/pii/000926149285330D}

\bibitem{2020SciPy-NMeth}
P.~Virtanen, R.~Gommers, T.~E. Oliphant, M.~Haberland, T.~Reddy, D.~Cournapeau,
  E.~Burovski, P.~Peterson, W.~Weckesser, J.~Bright, S.~J. {van der Walt},
  M.~Brett, J.~Wilson, K.~J. Millman, N.~Mayorov, A.~R.~J. Nelson, E.~Jones,
  R.~Kern, E.~Larson, C.~J. Carey, {\.I}.~Polat, Y.~Feng, E.~W. Moore,
  J.~{VanderPlas}, D.~Laxalde, J.~Perktold, R.~Cimrman, I.~Henriksen, E.~A.
  Quintero, C.~R. Harris, A.~M. Archibald, A.~H. Ribeiro, F.~Pedregosa, P.~{van
  Mulbregt}, {SciPy 1.0 Contributors},
  \href{https://www.nature.com/articles/s41592-019-0686-2}{{{SciPy} 1.0:
  Fundamental Algorithms for Scientific Computing in Python}}, Nat. Methods 17
  (2020) 261--272.
\newblock \href {https://doi.org/10.1038/s41592-019-0686-2}
  {\path{doi:10.1038/s41592-019-0686-2}}.
\newline\urlprefix\url{https://www.nature.com/articles/s41592-019-0686-2}

\bibitem{F2_III}
L.~Bytautas, N.~Matsunaga, T.~Nagata, M.~S. Gordon, K.~Ruedenberg,
  \href{https://doi.org/10.1063/1.2805392}{Accurate ab initio potential energy
  curve of {F}$_{2}$. {III}. {T}he vibration rotation spectrum}, J. Chem. Phys.
  127~(20) (2007) 204313.
\newblock \href {https://doi.org/10.1063/1.2805392}
  {\path{doi:10.1063/1.2805392}}.
\newline\urlprefix\url{https://doi.org/10.1063/1.2805392}

\bibitem{B2}
L.~Bytautas, N.~Matsunaga, G.~E. Scuseria, K.~Ruedenberg,
  \href{https://doi.org/10.1021/jp210473e}{Accurate potential energy curve for
  {B}$_{2}$. {A}b initio elucidation of the experimentally elusive ground state
  rotation-vibration spectrum}, J. Phys. Chem. A. 116~(7) (2012) 1717--1729.
\newblock \href {https://doi.org/10.1021/jp210473e}
  {\path{doi:10.1021/jp210473e}}.
\newline\urlprefix\url{https://doi.org/10.1021/jp210473e}

\bibitem{O2}
L.~Bytautas, N.~Matsunaga, K.~Ruedenberg,
  \href{https://doi.org/10.1063/1.3298376}{Accurate ab initio potential energy
  curve of {O}$_{2}$. {II}. {C}ore-valence correlations, relativistic
  contributions, and vibration-rotation spectrum}, J. Chem. Phys. 132~(7)
  (2010) 074307.
\newblock \href {https://doi.org/10.1063/1.3298376}
  {\path{doi:10.1063/1.3298376}}.
\newline\urlprefix\url{https://doi.org/10.1063/1.3298376}

\bibitem{PeFuncForDiatomicMolecules}
H.~M. Hulburt, J.~O. Hirschfelder,
  \href{https://doi.org/10.1063/1.1750827}{Potential energy functions for
  diatomic molecules}, J. Chem. Phys. 9~(1) (1941) 61--69.
\newblock \href {https://doi.org/10.1063/1.1750827}
  {\path{doi:10.1063/1.1750827}}.
\newline\urlprefix\url{https://doi.org/10.1063/1.1750827}

\bibitem{MpParamsForInteractions}
D.~D. Konowalow, J.~O. Hirschfelder,
  \href{https://aip.scitation.org/doi/abs/10.1063/1.1706374}{Morse potential
  parameters for {O}–{O}, {N}–{N}, and {N}–{O} interactions}, Phys.
  Fluids 4~(5) (1961) 637--642.
\newblock \href {https://doi.org/10.1063/1.1706374}
  {\path{doi:10.1063/1.1706374}}.
\newline\urlprefix\url{https://aip.scitation.org/doi/abs/10.1063/1.1706374}

\bibitem{ModelingOfDiatomicMolecule}
E.~Fidiani, \href{https://aip.scitation.org/doi/abs/10.1063/1.4943696}{Modeling
  of diatomic molecule using the {M}orse potential and the {V}erlet algorithm},
  AIP Conf. Proc. 1719~(1) (2016) 030001.
\newblock \href {https://doi.org/10.1063/1.4943696}
  {\path{doi:10.1063/1.4943696}}.
\newline\urlprefix\url{https://aip.scitation.org/doi/abs/10.1063/1.4943696}

\bibitem{F2_II}
L.~Bytautas, N.~Matsunaga, T.~Nagata, M.~S. Gordon, K.~Ruedenberg,
  \href{https://doi.org/10.1063/1.2801989}{Accurate ab initio potential energy
  curve of {F}$_{2}$. {II}. {C}ore-valence correlations, relativistic
  contributions, and long-range interactions}, J. Chem. Phys. 127~(20) (2007)
  204301.
\newblock \href {https://doi.org/10.1063/1.2801989}
  {\path{doi:10.1063/1.2801989}}.
\newline\urlprefix\url{https://doi.org/10.1063/1.2801989}

\bibitem{F2_Exp}
E.~A. Colbourn, M.~Dagenais, A.~E. Douglas, J.~W. Raymonda,
  \href{https://doi.org/10.1139/p76-159}{The electronic spectrum of {F}$_{2}$},
  Can. J. Phys. 54~(13) (1976) 1343--1359.
\newblock \href {https://doi.org/10.1139/p76-159} {\path{doi:10.1139/p76-159}}.
\newline\urlprefix\url{https://doi.org/10.1139/p76-159}

\bibitem{O2_Exp_0-22}
P.~H. Krupenie, \href{https://doi.org/10.1063/1.3253101}{The spectrum of
  molecular {O}xygen}, J. Phys. Chem. Ref. Data 1~(2) (1972) 423--534.
\newblock \href {https://doi.org/10.1063/1.3253101}
  {\path{doi:10.1063/1.3253101}}.
\newline\urlprefix\url{https://doi.org/10.1063/1.3253101}

\bibitem{O2_Exp_0-28}
D.~M. Creek, R.~W. Nicholls, A.~G. Gaydon,
  \href{https://royalsocietypublishing.org/doi/abs/10.1098/rspa.1975.0006}{A
  comprehensive re-analysis of the {O}$_{2}$ ({B}$^{3}{\Sigma}^{-}_{u}$ --
  {X}$^{3}{\Sigma}^{-}_{g})$ {S}chumann-{R}unge band system}, Proc. R. Soc. A:
  Math Phys. Sci. 341~(1627) (1975) 517--536.
\newblock \href {https://doi.org/10.1098/rspa.1975.0006}
  {\path{doi:10.1098/rspa.1975.0006}}.
\newline\urlprefix\url{https://royalsocietypublishing.org/doi/abs/10.1098/rspa.1975.0006}

\bibitem{O2_Exp_26-31}
R.~T. Jongma, S.~Shi, A.~M. Wodtke,
  \href{https://doi.org/10.1063/1.479618}{Electronic nonadiabaticity in highly
  vibrationally excited {O}$_{2}$ ({X}$^{3}{\Sigma}_{g}^{-})$: Spin-orbit
  coupling between {X}$^{3}{\Sigma}_{g}^{-}$ and b$^{1}{\Sigma}_{g}^{+}$}, J.
  Chem. Phys. 111~(6) (1999) 2588--2594.
\newblock \href {https://doi.org/10.1063/1.479618}
  {\path{doi:10.1063/1.479618}}.
\newline\urlprefix\url{https://doi.org/10.1063/1.479618}

\bibitem{O2_Exp_29-35}
X.~Yang, A.~M. Wodtke, \href{https://doi.org/10.1063/1.456240}{The potential
  energy function for {O}$_{2}$ ({X}$^{3}{\Sigma}_{g}^{-})$ and the transition
  dipole moment of the {S}chumann–{R}unge band near {X}‐state
  dissociation}, J. Chem. Phys. 90~(12) (1989) 7114--7117.
\newblock \href {https://doi.org/10.1063/1.456240}
  {\path{doi:10.1063/1.456240}}.
\newline\urlprefix\url{https://doi.org/10.1063/1.456240}

\bibitem{CanadianBoron}
A.~E. Douglas, G.~Herzberg,
  \href{https://doi.org/10.1139/cjr40a-016}{Spectroscopic evidence of the
  {B}$_2$ molecule and determination of its structure}, Can. J. Res. 18a~(11)
  (1940) 165--174.
\newblock \href {https://doi.org/10.1139/cjr40a-016}
  {\path{doi:10.1139/cjr40a-016}}.
\newline\urlprefix\url{https://doi.org/10.1139/cjr40a-016}

\bibitem{OldBoron}
H.~Bredohl, I.~Dubois, P.~Nzohabonayo,
  \href{https://www.sciencedirect.com/science/article/pii/0022285282901680}{The
  emission spectrum of {B}$_2$}, J. Mol. Spectrosc. 93~(2) (1982) 281--285.
\newblock \href {https://doi.org/https://doi.org/10.1016/0022-2852(82)90168-0}
  {\path{doi:https://doi.org/10.1016/0022-2852(82)90168-0}}.
\newline\urlprefix\url{https://www.sciencedirect.com/science/article/pii/0022285282901680}

\bibitem{NewBoron}
H.~Bredohl, I.~Dubois, F.~Melen, Excited vibrational levels of the ground state
  of {B}$_2$: The {A}$^3{\Sigma}^-_{u}$--{X}$^3{\Sigma}^-_{g}$ {T}ransition,
  Journal of Molecular Spectroscopy 121~(1) (1987) 128--134.

\end{thebibliography}

\end{document}